\documentstyle[12pt]{article}
\newcommand{\be}{\begin{eqnarray}}
\newcommand{\ee}{\end{eqnarray}}

\def\lsim{\mathrel{\rlap{\lower4pt\hbox{\hskip1pt$\sim$}}
    \raise1pt\hbox{$<$}}}                
\def\gsim{\mathrel{\rlap{\lower4pt\hbox{\hskip1pt$\sim$}}
    \raise1pt\hbox{$>$}}}                

\input{epsfig.sty}

\begin{document}

\Huge{\noindent{Istituto\\Nazionale\\Fisica\\Nucleare}}

\vspace{-3.9cm}

\Large{\rightline{Sezione SANIT\`{A}}}
\normalsize{}
\rightline{Istituto Superiore di Sanit\`{a}}
\rightline{Viale Regina Elena 299}
\rightline{I-00161 Roma, Italy}

\vspace{0.65cm}

\rightline{INFN-ISS 96/2}
\rightline{March 1996}

\vspace{1.5cm}

\begin{center}

\LARGE{Semi-inclusive Deep Inelastic Electron Scattering off the Deuteron and
the Neutron to Proton Structure Function Ratio \footnote{{\bf To appear in Phys.
Lett. B (1996).}}}\\

\vspace{1cm}

\large{Silvano Simula}

\vspace{0.5cm}

\normalsize{\em Istituto Nazionale di Fisica Nucleare, Sezione Sanit\`{a},\\
Viale Regina Elena 299, I-00161 Roma, Italy}

\end{center}

\vspace{1cm}

\begin{abstract}

The production of slow nucleons in semi-inclusive deep inelastic electron
scattering off the deuteron is investigated in the region $x \gsim 0.3$. It is
shown that within the spectator mechanism the semi-inclusive cross section
exhibits a scaling property even at moderate values of $Q^2$ ($\sim$ few
$(GeV/c)^2$) accessible at present facilities, like $CEBAF$. Such a scaling
property can be used as a model-independent test of the dominance of the
spectator mechanism itself and provides an interesting tool to investigate the
neutron structure function. The possibility of extracting model-independent
information on the neutron to proton structure function ratio from
semi-inclusive experiments is illustrated. The application of the spectator
scaling to semi-inclusive processes off complex nuclei is outlined.

\end{abstract}

\vspace{0.5cm}


\vspace{0.25cm}


\vspace{0.5cm}

\newpage

\pagestyle{plain}
 
\indent The investigation of deep inelastic scattering ($DIS$) of leptons off
the nucleon is an important tool to get fundamental information on the nucleon
wave function and the properties of its constituents. However, till now
experimental information on the structure function of the neutron has been
inferred from nuclear (usually deuteron) data by unfolding the neutron
contribution from the inclusive nuclear cross section. Such a procedure, which
typically involves the subtraction of both Fermi motion effects and
contributions from different nuclear constituents (i.e., nucleons, mesons,
isobars, ...), leads to non-trivial ambiguities related to the choice of the
model used to describe the structure of the target and the mechanism of the
reaction. As is well known, existing "experimental data" \cite{DATA,SLAC} on the
neutron structure function $F_2^n(x, Q^2)$ (where $x \equiv Q^2/ 2M\nu$ is the
Bjorken scaling variable and $Q^2 = - q^2  = |\vec{q}|^2 - {\nu}^2 > 0$ is the
squared four-momentum transfer) have been determined by combining deuteron,
$F_2^D(x, Q^2)$, and proton, $F_2^p(x, Q^2)$, inclusive data. It turns out
(see, e.g., \cite{SLAC}) that at $x \lsim 0.65$ the values obtained for the
neutron to proton structure function ratio, $R^{(n/p)}(x, Q^2) \equiv F_2^n(x,
Q^2) / F_2^p(x, Q^2)$, are almost independent of the assumed (non-relativistic)
deuteron wave function, whereas at larger values of $x$ there is a significant
model dependence. In particular, at $x \simeq 0.85$ the sensitivity of the
ratio $R^{(n/p)}(x, Q^2)$ to different deuteron wave functions corresponding to
various realistic models of the nucleon-nucleon interaction, reaches $\sim 20
\%$ \cite{SLAC}. Furthermore, corrections to the usual convolution formula
arise when off-shell components of the deuteron wave function as well as
off-mass-shell dependence of the nucleon structure function are taken into
account \cite{OFF-SHELL}. A recent estimate of these corrections \cite{MT96}
suggests a ratio $R^{(n/p)}(x, Q^2)$ at $x \to 1$ significantly larger than the
notorious "$1/4$" limiting value.  

\indent  An alternative way to obtain information on the neutron structure
function could be the investigation of semi-inclusive $DIS$ reactions of
leptons off the deuteron $^2H(\ell, \ell' N)X$ (cf. \cite{FS81}). The aim of
this letter is to address few relevant questions concerning the semi-inclusive
$DIS$ process $^2H(\ell,\ell'N)X$ at moderate and large values of $x$ ($x \gsim
0.3$) within the so-called spectator mechanism, according to which, after
lepton interaction with a quark of a nucleon in the deuteron, the spectator
nucleon is emitted because of recoil and detected in coincidence with the
scattered lepton. It will be shown that the cross section corresponding to such
a mechanism exhibits a scaling property (the spectator-scaling) even at
moderate values of $Q^2$ ($\sim$ few $(GeV/c)^2$), which are accessible at
present facilities, like, e.g., $CEBAF$. Such a scaling property can be used as
a model-independent test of the dominance of the spectator mechanism itself and
provides an interesting tool to extract the neutron structure function from
semi-inclusive data. Moreover, in the spectator-scaling regime the neutron to
proton structure function ratio $R^{(n/p)}(x, Q^2)$ can be obtained directly
from the ratio of the semi-inclusive cross sections of the processes
$^2H(e,e'p)X$ and $^2H(e,e'n)X$. Finally, the generalization of the
spectator-scaling to semi-inclusive $DIS$ processes off complex nuclei of the
type $A(\ell,\ell' (A-1)_{gr} )X$, where, besides the scattered lepton, the
residual ($A-1$)-nucleon system in its ground state is detected in the final
state, will be outlined.

\indent In case of electron scattering, the semi-inclusive cross section of
the process $^2H(e,e'N)X$ reads as follows
 \be
    {d^4 \sigma \over dE_{e'} ~ d\Omega_{e'} ~ dE_2 ~ d\Omega_2} = 
    \sigma_{Mott} ~ p_2 ~ E_2 ~ \sum_i ~ V_i ~ W_i^D(x, Q^2; \vec{p}_2)
     \label{1} 
 \ee
where $E_e$ ($E_{e'}$) is the initial (final) energy of the electron; $i \equiv
\{L, T, LT, TT\}$ identifies the different types of semi-inclusive response
functions ($W_i^D$) of the deuteron, $\vec{p}_2$ is the momentum of the
detected nucleon and $E_2 = \sqrt{M^2 + p_2^2}$ its energy ($p_2 \equiv
|\vec{p}_2|$). In Eq. (\ref{1}) $V_i$ is a kinematical factor, given explicitly
by $V_L = Q^4 / |\vec{q}|^4$, $V_T = tg^2(\theta_{e'} / 2) + Q^2 / 2
|\vec{q}|^2$, $V_{LT} = (Q^2 / \sqrt{2} |\vec{q}|^2) ~ \sqrt{ tg^2(\theta_{e'}
/ 2) + Q^2 / |\vec{q}|^2}$ and $V_{TT} = Q^2 / 2 |\vec{q}|^2$, where
$\theta_{e'}$ is the electron scattering angle.

\indent Let us consider the spectator mechanism, according to which the virtual
photon is absorbed by a quark belonging to the nucleon $N_1$ in the deuteron
and the recoiling nucleon $N_2$ is emitted and detected in coincidence with the
scattered electron. Within the plane wave impulse approximation and assuming a
non-relativistic deuteron wave function, the semi-inclusive deuteron response
function $W_i^D$ is related to the structure function $W_{\alpha}^{N_1}$ of the
struck nucleon by
 \be
    W_i^D(x, Q^2; \vec{p}_2) = {M \over E_2} n^{(D)}(p_2) ~ \sum_{\alpha = 1,2}
    ~ C_i^{\alpha}(x, Q^2; \vec{p}_2) ~ W_{\alpha}^{N_1}(M_1^*, Q^2)
    \label{3}
 \ee
where $n^{(D)}$ is the (non-relativistic) nucleon momentum distribution in the
deuteron and $M_1^*$ is the invariant mass of the struck nucleon. Using the
energy and momentum conservations, $M_1^*$ is explicitly given by $M_1^* =
\sqrt{(\nu + M_D - E_2)^2 - (\vec{q} - \vec{p}_2)^2}$, where $M_D$ is the
deuteron mass. The coefficients $C_i^{\alpha}(x, Q^2; \vec{p}_2)$, appearing in
Eq. (\ref{3}), depend upon the structure of the off-shell nucleonic tensor,
$W_{\mu \nu}^{N_1, off}$, and different prescriptions for the latter exist in
the literature (cf. \cite{OSC90}). In this letter we are mainly interested in
the production of slow nucleons ($p_2 \lsim 0.3 ~ GeV/c)$ and, therefore,
off-shell effects (being of relativistic origin) are not expected to play a
relevant role. Let us consider for $W_{\mu \nu}^{N_1, off}$ the following choice
(see \cite{OSC90})
 \be
    W_{\mu \nu}^{N_1, off} = - W_1^{N_1}(M_1^*, Q^2) \left [ g_{\mu \nu} +
    {q_{\mu} q_{\nu} \over Q^2} \right ] + {W_2^{N_1} (M_1^*, Q^2) \over M^2}
    \tilde{k}_{1, \mu}^{off} ~ \tilde{k}_{1, \nu}^{off}
    \label{5}
 \ee
where $W_{\alpha}^{N_1}(M_1, Q^2)$ is the on-shell nucleon structure function
and $\tilde{k}_{1, \mu}^{off} \equiv k_{1, \mu}^{off} + q_{\mu} (k_1^{off} \cdot
q) / Q^2$. The off-shell nucleon four-momentum is given by $k_1^{off}
\equiv (k_{1, 0}^{off}, \vec{k}_1) = (M_D - E_2, -\vec{p}_2)$, where the latter
equality follows from energy and momentum conservations. In principle, the
structure functions $W_{\alpha}^{N_1}$ appearing in Eq. (\ref{5}) should depend
not only on $M_1^*$ and $Q^2$, but also on $k_1^2 \equiv (M_D - E_2)^2 - p_2^2$.
From Eq. (\ref{5}) one has
 \be
    C_L^1 = - {|\vec{q}|^2 \over Q^2} ~~~~
    C_L^2 & = & {|\vec{q}|^4 \over Q^4} \left [ {(M_D - E_2) |\vec{q}| +
    \nu  p_2 cos(\theta_2) \over M |\vec{q}|} \right ] ^2 \nonumber \\
    C_T^1 = 2 ~~~~~~~~~
    C_T^2 & = & \left ( {p_2 sin(\theta_2) \over M} \right ) ^2 \nonumber \\
    C_{LT}^1 = 0 ~~~~~~~
    C_{LT}^2 & = & {\sqrt{8} p_2 sin(\theta_2) \over M} {|\vec{q}|^2 \over
    Q^2} {(M_D - E_2) |\vec{q}| + \nu  p_2 cos(\theta_2) \over M |\vec{q}|} 
    ~ cos(\phi_2) \nonumber \\
    C_{TT}^1 = 0 ~~~~~~~
    C_{TT}^2 & = & {1 \over 2} \left ( { p_2 sin(\theta_2) \over M} \right )^2
    cos(2\phi_2)
    \label{6}
\ee    
where $\theta_2$ and $\phi_2$ are the azimuth and polar angles of $\vec{p}_2$
with respect to $\vec{q}$, respectively.

\indent The relevant quantity, which will be discussed in this letter, is
related to the semi-inclusive cross section (\ref{1}) by
 \be
     F^{(s.i.)}(x, Q^2; \vec{p}_2) \equiv {1 \over K} ~ {d^4 \sigma \over
     dE_{e'} ~ d\Omega_{e'} ~ dE_2 ~ d\Omega_2}
     \label{7}
 \ee
where $K$ is a trivial kinematical factor, given by $K \equiv {2 M x^2 E_e
E_{e'} \over \pi Q^2} ~ {4 \pi \alpha^2 \over Q^4} ~ \left [ 1 - y + {y^2 \over
2} + {Q^2 \over 4 E_e^2} \right ]$, with $y \equiv \nu / E_e$. Eq. (\ref{7})
can be expressed in terms of the structure function $F_2^{N_1}(x^*, Q^2)$ of the
struck nucleon as
 \be
    F^{(s.i.)}(x, Q^2; \vec{p}_2) = M ~ p_2 ~ n^{(D)}(p_2) ~ {F_2^{N_1}(x^*,
    Q^2) \over x^*} ~ D^{N_1}(x, Q^2; \vec{p}_2)
    \label{8}
 \ee
where $x^* \equiv Q^2 / 2M \nu^*$, $\nu^* \equiv (Q^2 + M_1^{*2} - M^2) / 2M$
and
 \be
    D^{N_1}(x, Q^2; \vec{p}_2) = {1 - y - Q^2 / 4 E_e^2 \over 1 - y + y^2 /
    2 + Q^2 / 4 E_e^2} {\nu^2 \over \nu^{*2}} ~ \sum_i ~ V_i ~ \left [ C_i^2
    + C_i^1 ~ {1 + \nu^{*2} / Q^2 \over 1 + R_{L/T}^{N_1}(x^*, Q^2)} \right ]
    \label{10}
 \ee
with $R_{L/T}^N$ being the ratio of the longitudinal to transverse cross section
off the nucleon. Using Eq. (\ref{6}) it can be easily checked that in the
Bjorken limit (i.e., for $Q^2 \to \infty$ at fixed values of $x$ and $y$), the
contributions from the interference terms ($i = LT$ and $i = TT$) are
identically vanishing. Moreover, after considering $R_{L/T}^{N_1} \to_{Bj} 0$,
the longitudinal term vanishes too and, finally, one has $D^{N_1} \to_{Bj} 1$,
which implies $F^{(s.i.)}(x, Q^2; \vec{p}_2) \to_{Bj} M ~ p_2 ~ n^{(D)}(p_2) ~
F_2^{N_1}(x^*) / x^*$, where $F_2^{N_1}(x^*)$ stands for the nucleon structure
function in the Bjorken limit (apart from logarithmic $QCD$ corrections). It
follows that in the Bjorken limit and at fixed values of $p_2$ the function
$F^{(s.i.)}$ does not depend separately upon $x$ and the nucleon detection
angle $\theta_2$, but only upon the variable $x^*$, which in the Bjorken limit
becomes $x^* \to_{Bj} x / (2 - z_2)$, with $z_2 = [E_2 - p_2 cos(\theta_2)] /
M$ being the light-cone momentum fraction of the detected nucleon. Note that
any $p_2$-dependent deformation of the nucleon structure function (like, e.g.,
possible $k_1^2$-dependence) does not produce violations of the spectator
scaling property of $F^{(s.i.)}$. In what follows, we will refer to the function
$F^{(sp)}(x^*, Q^2, p_2)$ and variable $x^*$, given explicitly by
 \be
    F^{(sp)}(x^*, Q^2, p_2) & \equiv & M ~ p_2 ~ n^{(D)}(p_2) ~ F_2^{N_1}(x^*,
    Q^2) ~ / ~ x^*
    \label{11}  \\
    x^* & \equiv & {Q^2 \over Q^2 + (\nu + M_D - E_2)^2 - (\vec{q} -
    \vec{p}_2)^2 - M^2}
    \label{12}
 \ee
as the spectator-scaling function and variable, respectively. The essence of
the spectator scaling relies on the fact that the variable $x^*$ implies
different electron and nucleon kinematical conditions (in $x$ and $\theta_2$),
which correspond to the same value of the invariant mass produced on the struck
nucleon. Therefore, the deuteron response will be the same only if the
spectator mechanism dominates. Indeed, the spectator scaling is a peculiar
feature of the spectator mechanism and, therefore, its experimental observation
represents a test of the dominance of the spectator mechanism itself.

\indent An obvious question is whether an approximate spectator-scaling holds at
moderate values of $Q^2$ ($\sim$ few $(GeV/c)^2$), i.e. in a range of values of
$Q^2$ which can be reached, e.g., at $CEBAF$. To this end, the semi-inclusive
cross section (\ref{1}) for the process $^2H(e,e'p)X$ has been calculated using
Eqs. (\ref{3},\ref{6}) and assuming $E_e = 6 ~ GeV$, $Q^2 = 4 ~ (GeV/c)^2$ and
$p_2 = 0.1, ~ 0.3, ~ 0.5 ~ GeV/c$. The Bjorken variable $x$ and the nucleon
detection angle $\theta_2$ have been varied in the range $0.35 \div 0.95$ and
$10^o \div 150^o$, respectively (for sake of simplicity, the polar angle
$\phi_2$ has been chosen equal to $0$). As for the nucleon structure function,
the parametrization of the $SLAC$ data of Ref. \cite{SLAC}, containing both the
$DIS$ and nucleon resonance contributions, has been adopted. The results of the
calculations are shown in Fig. 1 and compared with the spectator-scaling
function (\ref{11}). It can clearly be seen that the spectator scaling is
almost completely fulfilled at $p_2 \simeq 0.1 ~ GeV/c$, whereas
spectator-scaling violations are relevant already at $p_2 = 0.5 ~ GeV/c$,
preventing the nucleon resonance peaks to be observed. This result is due to
the $\vec{p}_2$-dependence of the coefficients $C_i^2$ (see Eq. (\ref{6})),
produced mainly by the convective part of the electromagnetic current (note that
at $\vec{p}_2 = 0$ one has $D^{N_1}(x, Q^2; \vec{p}_2 = 0) = 1$). The results
obtained for the quantity $D^n(x, Q^2; \vec{p}_2)$ (Eq. (\ref{10})), which
represents the ratio $F^{(s.i.)}(x, Q^2, \vec{p}_2) / F^{(sp)}(x^*, Q^2, p_2)$,
are reported in Figs. 2a and 2b for forward ($\theta_2 < 90^o$) and backward
($\theta_2 > 90^o$) proton emission, respectively. Moreover, the effects on
$D^n(x, Q^2; \vec{p}_2)$ due to different off-shell prescriptions for the
nucleonic tensor $W_{\mu \nu}^{N_1, off}$ have been investigated. In Fig. 2 the
open squares are the results of the calculations performed following the
prescription of Ref. \cite{HT90}, according to which in the nucleonic tensor
(\ref{5}) $k_1^{off}$ is replaced by $k_1^{on} \equiv (\sqrt{M^2 +
|\vec{k}_1|^2}, \vec{k}_1) = (E_2, -\vec{p}_2)$ and the four-momentum transfer
$q = (\nu, \vec{q})$ by $\bar{q} = (\bar{\nu}, \vec{q})$ with $\bar{\nu} = \nu
+ k_{1,0}^{off} - k_{1,0}^{on} = M_D - 2 E_2$. The following comments are in
order: i) the limiting value $D^n = 1$ is reached within $\sim 20 \%$; ii)
spectator scaling violations at $Q^2 = 4 ~ (GeV/c)^2$ are $\sim 15 \%$ in case
of forward proton emission and only within $\sim 5 \%$ for backward proton
kinematics, thanks to the $cos(\theta_2)$-dependence of the coefficients $C_L^2$
and $C_{LT}^2$; iii) at low values of $p_2$ ($\lsim 0.3 ~ GeV/c$) the effects
due to different off-shell prescriptions for the nucleonic tensor turns out to
be quite small (cf. also \cite{UKK96}).   

\indent In the spectator-scaling regime the measurement of the semi-inclusive
cross section both for $^2H(e,e'p)X$ and $^2H(e,e'n)X$ processes would allow
the investigation of two spectator-scaling functions, involving the same
nuclear part, $M p_2 n^{(D)}(p_2)$, and the neutron and proton structure
functions, respectively. Therefore, assuming $R_{L/T}^n = R_{L/T}^p$ (as it is
suggested by recent $SLAC$ data analyses \cite{RLT_SLAC}), both the nuclear part
and the factor $D^{N_1}(x, Q^2; \vec{p}_2)$ cancel out in the ratio of the
semi-inclusive cross sections $R^{(s.i.)}(x, Q^2, \vec{p}_2) \equiv d^4
\sigma[^2H(e,e'p)X] /$ $d^4 \sigma[^2H(e,e'n)X]$, which provides in this way
directly the neutron to proton structure function ratio $R^{(n/p)}(x^*, Q^2)$.
It follows that, with respect to the function $F^{(s.i.)}(x, Q^2, \vec{p}_2)$,
the ratio $R^{(s.i.)}(x, Q^2, \vec{p}_2)$ exhibits a more general scaling
property, for at fixed $Q^2$ it does not depend separately upon $x$, $p_2$ and
$\theta_2$, but only on $x^*$. This means that any $p_2$-dependence of the
ratio $R^{(s.i.)}$ would allow to investigate off-shell deformations of the
nucleon structure functions (see below).

\indent The results presented and, in particular, the spectator-scaling
properties of $F^{(s.i.)}$ and $R^{(s.i.)}$ could in principle be modified by
the effects of mechanisms different from the spectator one, like, e.g., the
fragmentation of the struck nucleon, or by the breakdown of the impulse
approximation (\ref{3}). In order to estimate the effects of the so-called
target fragmentation of the struck nucleon (which is thought to be responsible
for the production of slow hadrons in $DIS$ processes), we make use of the same
approach already applied in Ref. \cite{SIM93} for investigating the nucleon
emission in semi-inclusive $DIS$ of leptons off light and complex nuclei. In
\cite{SIM93} the hadronization mechanism is parametrized through the use of
fragmentation functions, whose explicit form has been chosen according to the
prescription of Ref. \cite{BDS92}, elaborated to describe the production of
slow protons in $DIS$ of (anti)neutrinos off hydrogen and deuterium targets.
Furthermore, the effects arising from possible six-quark ($6q$) cluster
configurations at short internucleon separations, are explicitly considered.
According to the mechanism first proposed in Ref. \cite{SIX-QUARKS}, after
lepton interaction with a quark belonging to a $6q$ cluster, nucleons can be
formed out of the penta-quark residuum and emitted forward as well as backward.
The details of the calculations can be easily inferred from Ref. \cite{SIM95},
where $6q$ bag effects in semi-inclusive $DIS$ of leptons off light and complex
nuclei have been investigated. The estimate of the nucleon production, arising
from the above-mentioned target fragmentation processes, is shown in Fig. 3 for
the function $F^{(s.i.)}$ and in Fig. 4 for the ratio $R^{(s.i.)}$. It can
clearly be seen that: i) only at $x^* \lsim 0.4$ the fragmentation processes can
produce relevant violations of the spectator scaling (see Figs. 3 and 4(a)); ii)
backward kinematics (see Fig. 4(b)) appear to be the most appropriate conditions
to extract the neutron to proton ratio $R^{(n/p)}$. These results are not
unexpected, because ~ i) fragmentation processes produce only forward nucleons
in a frame where the struck nucleon is initially at rest, and ~ ii) target
fragmentation is mainly associated with a diquark remnant carrying a light-cone
momentum fraction $\sim (1 - x^*)$, which vanishes as $x^* \to 1$ (cf. also
\cite{SIM93,SIM95}). Moreover, explicit calculations show that the relevance of
the fragmentation processes drastically decreases when $p_2 < 0.5 ~ GeV/c$.

\indent As far as the impulse approximation is concerned, it should be reminded
that our calculations have been performed within the assumption that the debris
produced by the fragmentation of the struck nucleon does not interact with the
recoiling spectator nucleon. Estimates of the final state interactions of the
fragments in semi-inclusive processes off the deuteron have been obtained in
\cite{TN92}, suggesting that rescattering effects should play a minor role
thanks to the finite formation time of the dressed hadrons. Moreover, backward
nucleon emission is not expected to be sensitively affected by forward-produced
hadrons (see \cite{BDT94}).

\indent Besides fragmentation and final state interaction, also nucleon
off-shell effects might produce violations of the spectator scaling, in
particular at high values of $p_2$ ($\gsim 0.3 ~ GeV/c$). The results of the
calculations of the ratio $R^{(s.i.)}(x, Q^2, \vec{p}_2)$, obtained considering
the off-shell effects suggested in Refs. \cite{DT86}\footnote{In \cite{DT86} a
$Q^2$-rescaling model, based on a rescaling parameter related to the virtuality
$k_1^2 = (M_D - E_2)^2 - p_2^2$ of the struck nucleon, is applied to the
nucleon structure functions appearing in Eq. (\ref{5}).} and \cite{HT90}, are
shown in Fig. 5(a) and 5(b), respectively. It can be concluded that the
measurement of the ratio $R^{(s.i.)}(x, Q^2, \vec{p}_2)$ represents an
interesting tool both to investigate the ratio of free neutron to proton
structure function, provided $p_2 \sim 0.1 \div 0.2 ~ GeV/c$, and to get
information on possible off-shell behaviour of the nucleon structure function
when $p_2 \gsim 0.3 ~ GeV/c$.

\indent Before closing, we want to address briefly the question whether
semi-inclusive processes off complex nuclei exhibit the spectator-scaling
property. A detailed analysis is in progress; here, we want only to mention that
the results previously obtained in case of the process $^2H(e,e'N)X$ can be
easily generalized to reactions of the type $A(e,e' (A-1)_{gr})X$, in which,
besides the scattered electron, the residual ($A-1$)-nucleon system in its
ground state (or in any state belonging to its discrete spectrum) is detected
in the final state. For these processes the spectator-scaling function and
variable are explicitly given by
 \be
    F^{(sp)}(x^*, Q^2, k_{A-1}) & = & {M ~ k_{A-1} ~ E_{A-1} ~
    n^{(gr)}(k_{A-1}) \over \sqrt{M^2 + k_{A-1}^2}} ~ {F_2^N(x^*, Q^2) \over
    x^*} \nonumber \\
    x^* & \equiv & {Q^2 \over Q^2 + (\nu + M_A - \sqrt{M_{A-1}^2 +
    k_{A-1}^2})^2 - (\vec{q} - \vec{k}_{A-1})^2 - M^2}
    \label{13}
 \ee     
where $N$ is the missing nucleon, $\vec{k}_{A-1}$ the momentum of the detected
nucleus ($A-1$), $E_{A-1} = \sqrt{M_{A-1}^2 + k_{A-1}^2}$ its total energy and
$n^{(gr)}(k)$ the nucleon momentum distribution corresponding to the
ground-to-ground transition. Note that, in case of $A > 2$, no nucleon
fragmentation process can be in competition with the spectator mechanism.

\indent In conclusion, the production of slow nucleons in semi-inclusive deep
inelastic electron scattering off the deuteron has been investigated in the
kinematical regions corresponding to $x \gsim 0.3$. It has been shown that
within the spectator mechanism the semi-inclusive cross section exhibits an
interesting scaling property even at moderate values of $Q^2$ ($\sim$ few
$(GeV/c)^2$) accessible at present facilities, like, e.g., $CEBAF$. It has been
pointed out that the spectator scaling can be used as a model-independent test
of the dominance of the spectator mechanism itself and allows the investigation
of the neutron structure function from semi-inclusive data. In the
spectator-scaling regime the neutron to proton structure function ratio can be
obtained directly from the ratio of the semi-inclusive cross sections of the
processes $^2H(e,e'p)X$ and $^2H(e,e'n)X$. Finally, the spectator-scaling
property can be easily generalized to semi-inclusive processes off complex
nuclei of the type $A(e,e' (A-1)_{gr} )X$, where the residual ($A-1$)-nucleon
system in its ground state is detected in the final state.

\newpage

\begin{center}

{\bf Figure Captions}

\end{center}

\vspace{0.10cm}

Fig. 1. (a) The function $F^{(s.i.)}(x, Q^2, \vec{p}_2)$ (Eq. (\ref{7}))
for the process $^2H(e,e'p)X$ plotted versus the spectator-scaling variable
$x^*$ (Eq. (\ref{12})) at $Q^2 = 4 ~ (GeV/c)^2$ and $p_2 = 0.1 ~ GeV/c$. The
calculation of the semi-inlcusive cross section (\ref{1}) is based on the
spectator mechanism (see Eqs. (\ref{3},\ref{6})), assuming the prescription
(\ref{5}) for the off-shell nucleonic tensor. The values of $x$ have been
varied in the range $0.35 \div 0.95$. The open (full) dots, squares, diamonds
and triangles correspond to  $\theta_2 = 10^o$ ($100^o$), $30^o$ ($110^o$),
$50^o$ ($130^o$), $70^o$ ($150^o$), respectively. The solid line is the
spectator-scaling function $F^{(sp)}(x^*, Q^2, p_2)$ (Eq. (\ref{11}))
calculated using the deuteron momentum distribution corresponding to the Paris
nucleon-nucleon interaction \cite{PARIS} and to the parametrization of the
neutron structure function of Ref. \cite{SLAC}. (b) The same as in (a), but at
$p_2 = 0.5 ~ GeV/c$.

\vspace{0.10cm}

Fig. 2. The quantity $D^n(x, Q^2, \vec{p}_2)$ (Eq. (\ref{10})) for the
process $^2H(e,e'p)X$ plotted versus the spectator-scaling variable $x^*$ (Eq.
(\ref{12})) at $Q^2 = 4 ~ (GeV/c)^2$ and $p_2 = 0.1 ~ GeV/c$. The values of $x$
and $\theta_2$ are the same as in Fig. 1. Forward ($\theta_2 < 90^o$) and
backward ($\theta_2 > 90^o$) proton emission are shown in (a) and (b),
respectively. The full dots correspond to the results of the calculations based
on Eq. (\ref{5}) for the off-shell nucleonic tensor, whereas the open squares
are the results obtained following the off-shell prescription of Ref.
\cite{HT90}. 

\vspace{0.10cm}

Fig. 3. The quantity $F^{(s.i.)}(x, Q^2, \vec{p}_2)$ (Eq. (\ref{7})) for the
process $^2H(e,e'p)X$ plotted versus the spectator-scaling variable $x^*$ (Eq.
(\ref{12})) at $Q^2 = 4 ~ (GeV/c)^2$ and $p_2 = 0.5 ~ GeV/c$. The calculations
include the effects of the target fragmentation of the struck nucleon,
evaluated as in Ref. \cite{SIM93}, as well as the contribution of the proton
emission arising from virtual photon absorption on a $6q$ cluster configuration
in the deuteron, evaluated following Ref. \cite{SIM95} and adopting a $6q$ bag
probability equal to $2 \%$. The values of $x$ and $\theta_2$, as well as the
solid line, are the same as in Fig. 1(b).

\vspace{0.10cm}

Fig. 4. (a) The ratio $R^{(s.i.)}(x, Q^2, \vec{p}_2)$ of the semi-inclusive
cross sections (\ref{1}) for the processes $^2H(e,e'p)X$ and $^2H(e,e'n)X$,
calculated at $Q^2 = 4 ~ (GeV/c)^2$ and $x$ in the range $0.35 \div 0.95$. The
full (open) dots, squares, diamonds and triangles correspond to $p_2 = 0.3 ~
(0.5) ~ GeV/c$ and $\theta_2 = 10^o, 30^o, 50^o, 70^o$, respectively. The
results obtained at $p_2 = 0.1 ~ GeV/c$ and $\theta_2 = 10^o, 30^o, 50^o, 70^o$
are represented by the crosses, plus signs, full and open (downward) triangles,
respectively. The solid line is the neutron to proton structure function ratio
$R^{(n/p)}(x^*, Q^2)$ calculated using the parametrization of the nucleon
structure function of Ref. \cite{SLAC}, which predicts a (model-dependent)
limiting value of $\sim 0.35$ at $x^* \to 1$ and $Q^2 = 4 ~ (GeV/c)^2$. (b) The
same as in (a), but for backward nucleon emission at $\theta_2 = 100^o, 110^o,
130^o, 150^o$.

\vspace{0.10cm}

Fig. 5. The ratio $R^{(s.i.)}(x, Q^2, \vec{p}_2)$ of the semi-inclusive cross
sections for the processes $^2H(e,e'p)X$ and $^2H(e,e'n)X$, calculated
considering the off-shell effects proposed in Refs. \cite{DT86} (a) and
\cite{HT90} (b), respectively. Backward nucleon emission only ($\theta_2 >
90^o$) has been considered. The dots, squares and triangles correspond to $p_2
= 0.1, ~ 0.3, ~ 0.5 ~ GeV/c$, respectively. The solid line is the same as in
Fig. 4.

\newpage

\begin{figure}

\epsfig{file=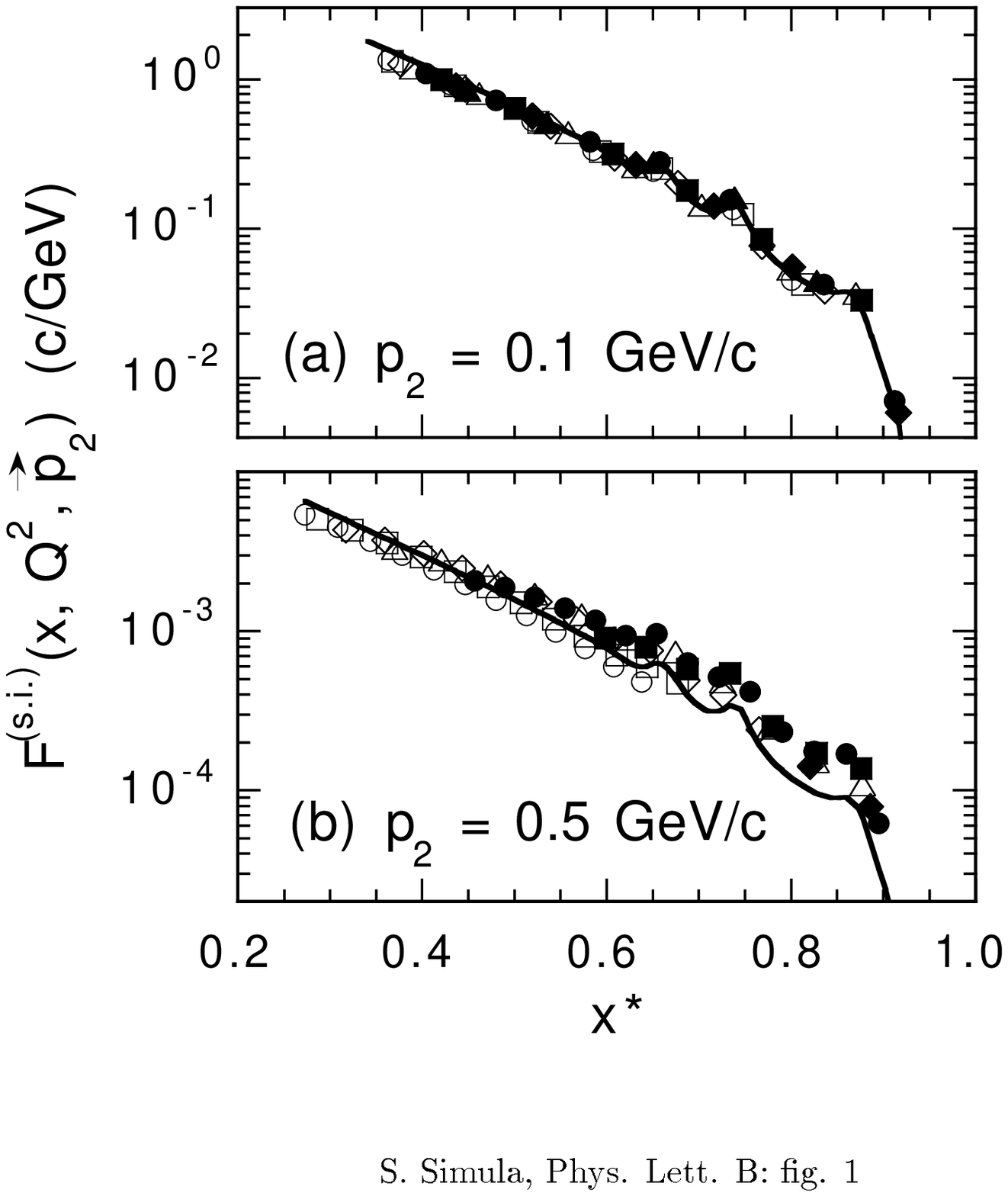}

\end{figure}

\newpage

\begin{figure}

\epsfig{file=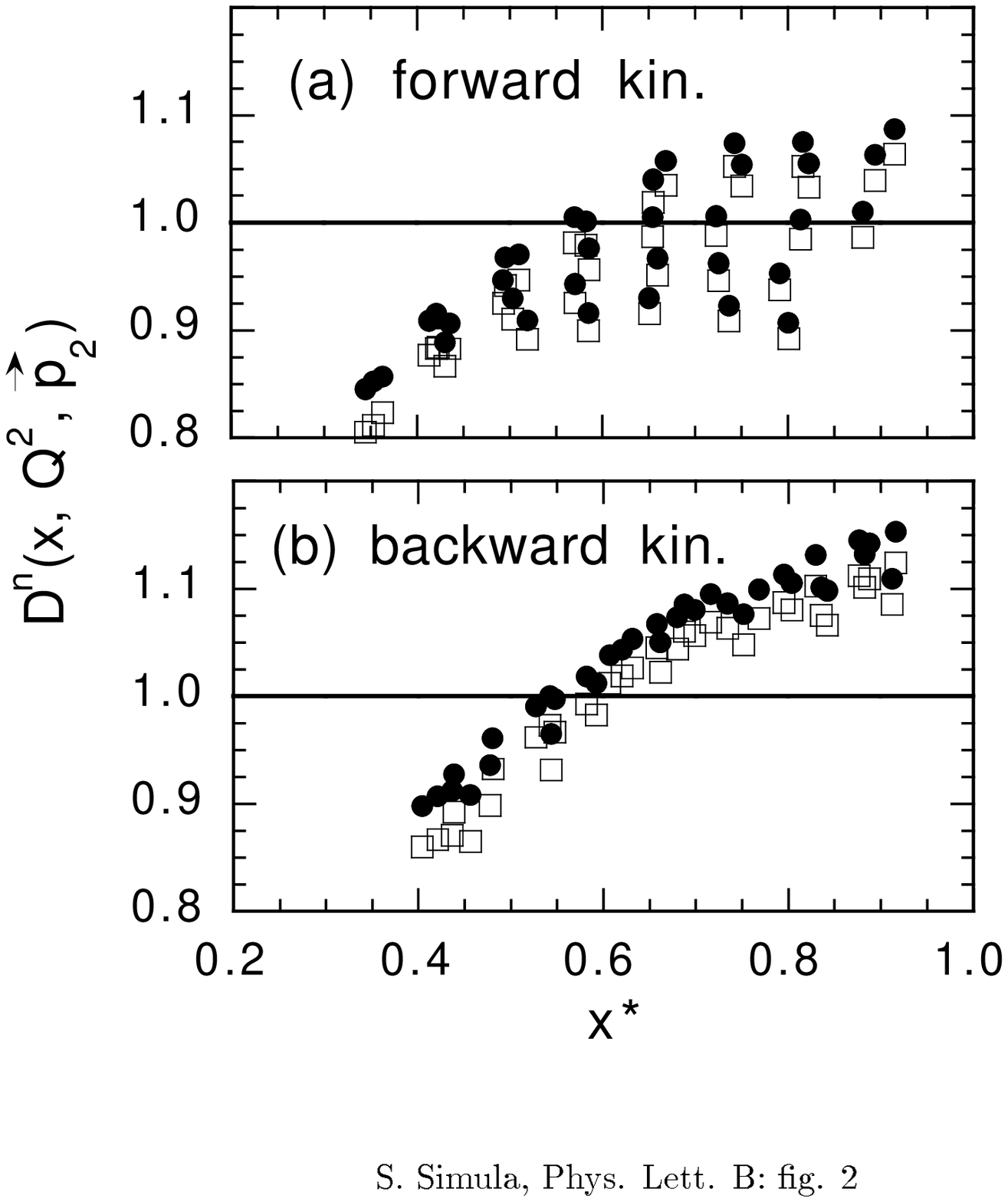}

\end{figure}

\newpage

\begin{figure}

\epsfig{file=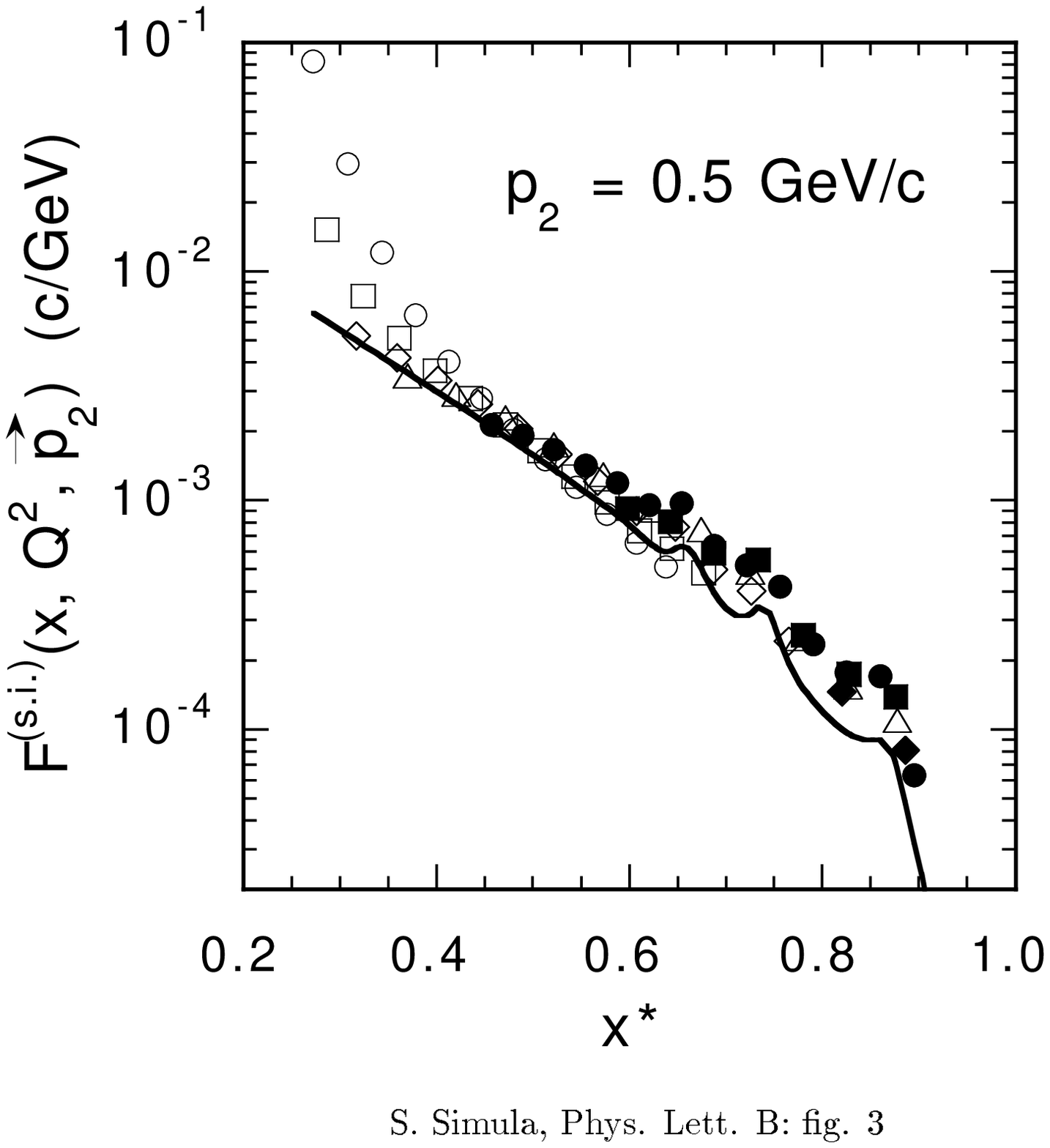}

\end{figure}

\newpage

\begin{figure}

\epsfig{file=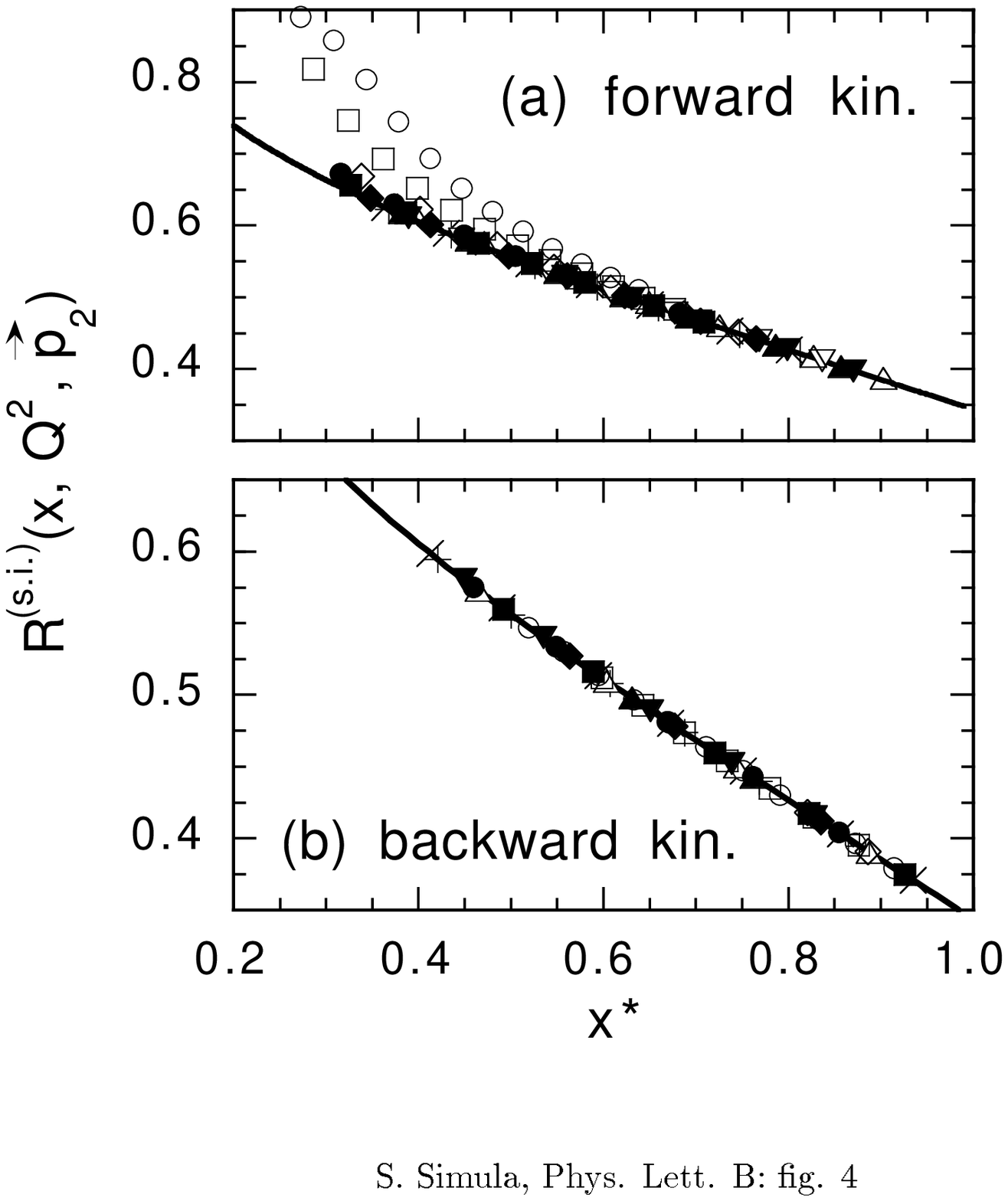}

\end{figure}

\newpage

\begin{figure}

\epsfig{file=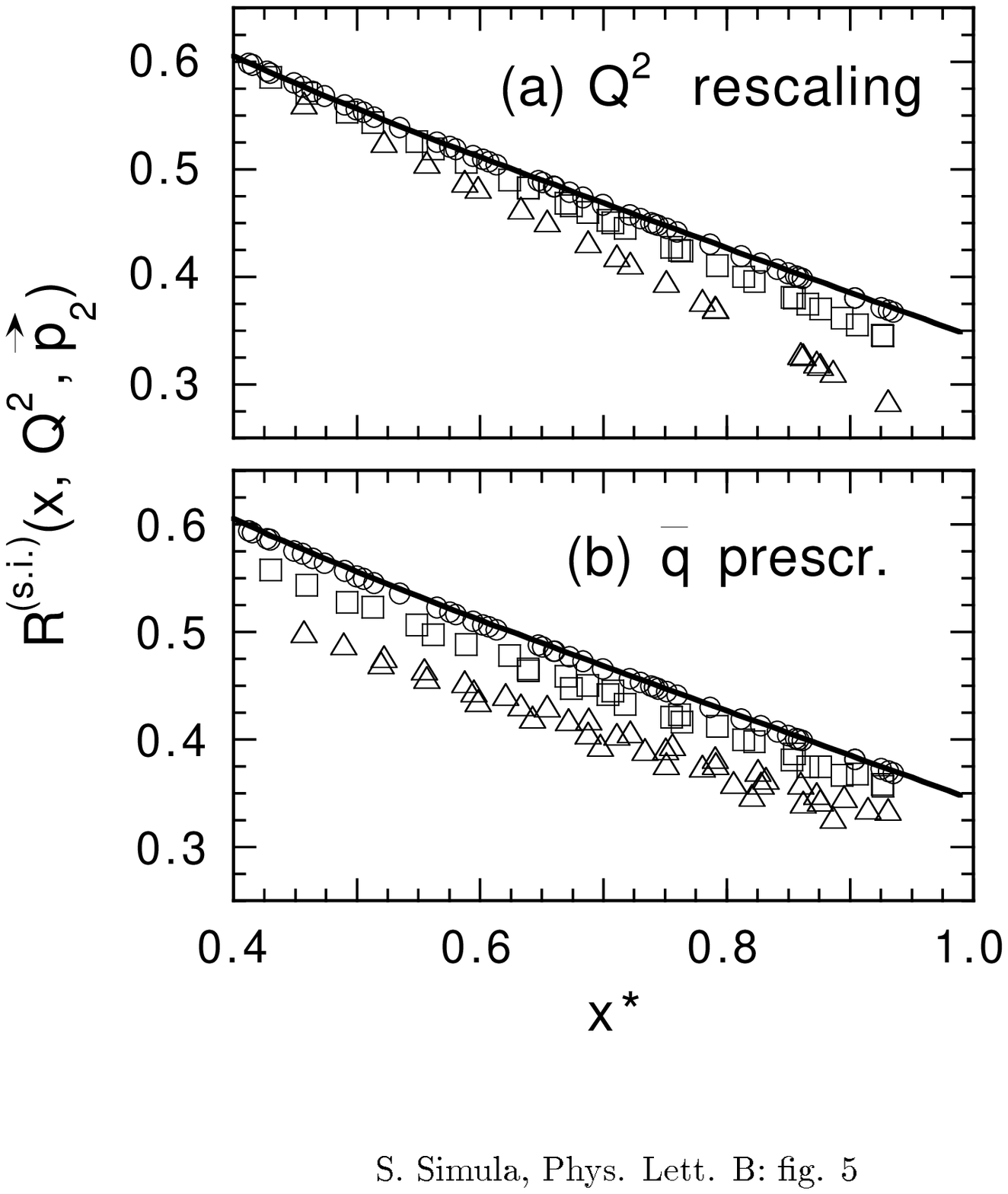}

\end{figure}

\end{document}